\documentclass[manuscript]{acmart}

\AtBeginDocument{%
  \providecommand\BibTeX{{%
    \normalfont B\kern-0.5em{\scshape i\kern-0.25em b}\kern-0.8em\TeX}}}

\copyrightyear{2021}
\acmYear{2021}
\setcopyright{acmcopyright}\acmConference[IUI '21]{26th International Conference on Intelligent User Interfaces}{April 14--17, 2021}{College Station, TX, USA}
\acmBooktitle{26th International Conference on Intelligent User Interfaces (IUI '21), April 14--17, 2021, College Station, TX, USA}
\acmPrice{15.00}
\acmDOI{10.1145/3397481.3450696}
\acmISBN{978-1-4503-8017-1/21/04}

\acmConference[IUI '21]{IUI '21: ACM IUI 2021}{April 13--17, 2021}{College Station, Texas}
\usepackage{CJKutf8}

\begin{document}

\title{Improving Artificial Teachers by Considering How People Learn and Forget}

\author{Aur\'elien Nioche}
\authornote{Corresponding author.}
\orcid{0000-0002-0567-2637}
\affiliation{%
  \institution{Aalto University}
  \city{Helsinki}
  \country{Finland}}
 \email{nioche.aurelien@gmail.com}

\author{Pierre-Alexandre Murena}
\orcid{0000-0003-4586-9511}
\affiliation{%
  \institution{Aalto University}
  \city{Helsinki}
  \country{Finland}}
\email{pierre-alexandre.murena@aalto.fi}

\author{Carlos de la Torre-Ortiz}
\orcid{0000-0002-7457-5216}
\affiliation{%
  \institution{Aalto University}
  \city{Helsinki}
  \country{Finland}
 }
 \affiliation{%
  \institution{University of Helsinki}
  \city{Helsinki}
  \country{Finland}
 }
\email{carlos.delatorreortiz@helsinki.fi}

\author{Antti Oulasvirta}
\affiliation{%
  \institution{Aalto University}
  \city{Helsinki}
  \country{Finland}
 }
\email{antti.oulasvirta@aalto.fi}

\renewcommand{\shortauthors}{Nioche et al.}

\begin{abstract}
The paper presents a novel model-based method for intelligent tutoring, with particular emphasis on the problem of selecting teaching interventions in interaction with humans. Whereas previous work has focused on either personalization of teaching or optimization of teaching intervention sequences, the proposed individualized model-based planning approach represents convergence of these two lines of research. Model-based planning picks the best interventions via interactive learning of a user memory model's parameters. The approach is novel in its use of a cognitive model that can account for several key individual- and material-specific characteristics related to recall/forgetting, along with a planning technique that considers users' practice schedules. Taking a rule-based approach as a baseline, the authors evaluated the method's benefits in a controlled study of artificial teaching in second-language vocabulary learning ($N=53$).

\end{abstract}


% A "teaser" image appears between the author and affiliation
% information and the body of the document, and typically spans the
% page.
\begin{teaserfigure}
  \centering
  \includegraphics[width=\textwidth]{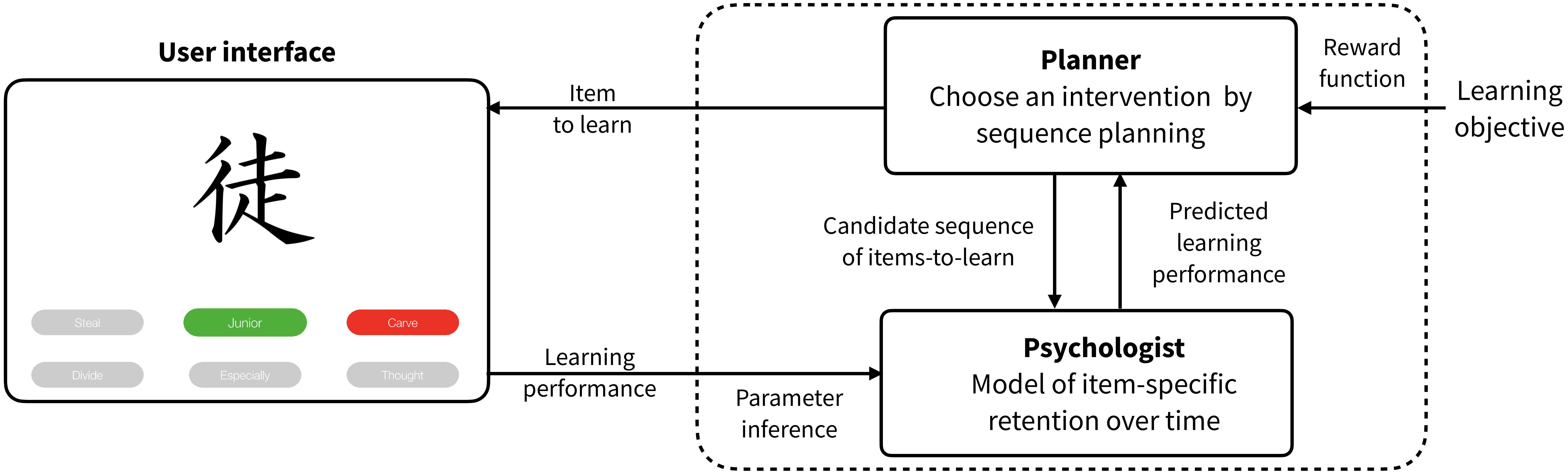}
  \caption{
     We propose a model-based planning approach for intelligent tutoring that combines parameter inference and sequence planning.
     Informed by user performance, the psychologist component updates its beliefs about each user's memory state for each item. To predict the consequences of particular candidate sequences for the human learner, the planner component consults the psychologist; then, the planner chooses the teaching intervention that maximizes the potential for meeting the learning objective.
  }
  \Description{Description of the model's architecture:
  At the left is a screenshot of the graphical user interface (GUI) of our Web-based application.
  On the right is a diagram of our framework.
  The GUI presents a kanji (a logographic Chinese character in the Japanese alphabet).
  Six possible answers are displayed at the bottom. If the user selects an incorrect one, it gets displayed in red and the correct one is shown in green.
  The diagram shows that a model-based artificial teacher comprises a planner and a psychologist.
  This system is supplied with an external learning objective, with an associated reward function for the planner.
  The planner's role is to choose an intervention.
  The psychologist is in charge of the model for item-specific retention in memory over time.
  It and the planner engage in interplay wherein the planner provides the psychologist with a candidate sequence of items to learn and the psychologist presents the predicted learning performance.
  The GUI gives the psychologist input on actual learning performance, which enables parameter inference.
  The planner's final output feeds back to the GUI, supplying an item to be learned.
  }\label{fig:teaser}
\end{teaserfigure}

%
% This command processes the author and affiliation and title
% information and builds the first part of the formatted document.
\maketitle

\section{Introduction}
\textit{Intelligent tutoring} \cite{anderson1985intelligent} addresses the problem of designing teaching interventions -- any materials, practice, and feedback delivered by a teacher -- for education objectives such as learning a new language or skill.
Applications for self-regulated teaching are very popular (e.g.,\ with Duolingo estimates of 100M+ downloads from Google Play at the time of writing).
One of the central challenges for research on intelligent user interfaces is to identify algorithmic principles that can pick the best interventions for reliably improving human learning toward stated objectives in light of realistically obtainable data on the user.
The computational problem we study is how, when given some learning materials, we can organize them into lessons and reviews such that, over time, human learning is maximized with respect to a set learning objective.

Predicting the effects of teaching interventions on human learning is challenging, however.
Firstly, the state of user memory is both \textit{latent} (that is, not directly observable) and \textit{non-stationary} (that is, evolving over time, on account of such effects as loss of activation and interference), and an intervention that is ideal for one user may be a poor choice for another user --- there are large individual-to-individual differences in forgetting and recall.
Secondly, planning in this context is particularly difficult: the number of possible sequences is generally large, and the reward is often delayed. Rarely is the goal to maximize knowledge at the time of learning, as opposed to some later stage/point (e.g.\ an exam date).

Previous work on this topic has mainly explored rule-based and model-based approaches (see ``Related Work'').
One popular rule-based method is the Leitner system.
Following the principle that spacing out the reviews of an item is beneficial for the teaching process~\cite{ebbinghaus1885uber}, it lowers the frequency of review when the student has responded correctly and otherwise increases it~\cite{pimsleur1967memory,leitner1972so,wozniak1994optimization}.
Although this approach is easy to implement, computationally inexpensive, and applicable across different contexts~\cite{gerbier2015effect,kang2016spaced} and while there is evidence that it yields benefits in terms of learning optimization~\cite{reddy2016unbounded}, its ability to adapt to the individual's characteristics is limited.
It considers only successes and errors, not addressing individual-specific characteristics such as the user's forgetting rate or time constraints. Model-based approaches, on the other hand, utilize a predictive model of (human) memory to select interventions or schedule practice sessions~\cite{pavlik2008using,lindsey2014improving,settles2016trainable,tabibian2019enhancing,hunziker2019teaching,aydin2020broccoli}.
While results have been positive for models that represent population-level characteristics, these still do not model each user individually (i.e.\ a particular user's memory state is not inferred).
Indeed, fitting a memory model to the individual's learning data in \textit{online} fashion (i.e.\ during interaction with the user) is rendered challenging by the scarcity of data and the typically large number of parameters in these models.
Also, instead of planning a sequence of actions, most studies either used a heuristic approach for selection of the next item~\cite{lindsey2014improving,hunziker2019teaching} or applied \textit{offline} optimization~\cite{settles2016trainable,tabibian2019enhancing,aydin2020broccoli}, hence limiting adaptability and real-world applicability.

With this paper, we examine a \textbf{model-based planning} approach that holds potential for better adapting to individuals' learning characteristics.
It contributes to scholarship by combining planning and inference with model-based approaches at the level of individual learners and items. More specifically, proceeding from the seminal work of Rafferty et al.~\cite{rafferty2016faster}, we frame the item-selection process as a partially observable Markov decision process (POMDP) and introduce a modular framework for (i) inference of memory state at the item level for each user and (ii) selection of the next item as planning a sequence of interventions. We assess the performance of a simple implementation of our framework, benchmarking it against a well-established rule-based baseline (the aforementioned Leitner system, which is ubiquitous in commercial deployments of self-regulated teaching applications). We consider both artificial agents and a controlled study conducted with human learners over one week's training ($N=53$).

\section{Related Work}
\subsection{Rule-based systems}

Rule-based approaches define hand-crafted rules for deciding which learning events to trigger on the basis of the response data from the user.
One of the first implementations applied an approach suggested by Pimsleur~\shortcite{pimsleur1967memory} in which a rigid schedule progressively spaces out the reviews in line with a power law.
A few years later, Leitner~\shortcite{leitner1972so} proposed a simple rule for adapting to the user: in conditions of success, lower the frequency of review; in those of failure, increase it.
Atkinson~\shortcite{atkinson1972optimizing} proposed one implementation of this algorithm,
a popular variant of which is still widely used today, under the name \textit{SuperMemo}~\cite{wozniak1994optimization}.

\subsection{Model-free and model-based learning}

The problem of intelligent tutoring can also be formulated as that of maximizing learning over a sequence of learning events.
In so-called model-free approaches, a \textit{teaching policy} is learned via experience.
For instance, Clement and colleagues~\shortcite{clement2015multi} proposed framing the teaching situation as a multi-armed bandit system.
The downside of this approach is the same as that of model-free methods in general: they require extensive training and generalize poorly to unseen situations.
Model-based approaches, in contrast, rely on a predictive model of human memory for scheduling the reviews.

\paragraph{Myopic planning.} Lindsey et al.~\shortcite{lindsey2014improving} took a model-based approach to optimize scheduling of reviews, by relying on ACT-R as the memory model~\cite{anderson1996act,pavlik2003act,pavlik2005practice}.
No planning is implemented in this technique, though, because the intelligent tutoring system uses a heuristic for item selection:
it employs a threshold $\tau$ such that it presents the user with any item for which the predicted probability of recall falls below the threshold set.
More recently, working with an exponential-decay model of memory~\cite{ebbinghaus1885uber}, Hunziker et al.~\shortcite{hunziker2019teaching} also used a myopic planner.
For their context, they showed that a myopic planner should perform at least as well as a non-myopic one.
However, their context does not assume sessions separated by long breaks, so the scope for application of their results is limited.

\paragraph{Offline optimization.} Pavlik et al.~\shortcite{pavlik2008using} too based their intelligent tutoring system on ACT-R~\cite{anderson1996act,pavlik2003act,pavlik2005practice}, but they provided a non-myopic planning technique in addition.
Still, the model's parameters that support the planning are optimized offline, kept constant, and not fitted to the current user's data.
More recently, Settles et al.~\shortcite{settles2016trainable}, Tabibian et al.~\shortcite{tabibian2019enhancing}, and Aydin et al.~\shortcite{aydin2020broccoli} proposed fitting a parameterizable memory model to user data.
Their approaches build on an exponential forgetting model to infer a distribution of probability of recall for the items and then, given this distribution, select the next item to present.
Unlike our method, these two models nonetheless assume that the best-fitting parameter values are common to the entire population, and the optimization is done offline.

\paragraph{Planning as a POMDP} Rafferty et al.~\shortcite{rafferty2016faster} have framed the teaching situation as a POMDP, to enable dealing with the teacher's uncertainty about the cognitive state of the learner (the literature discusses various POMDP-based models of human behavior~\cite{howes2018interaction}).
Whitehill et al.~\shortcite{whitehill2017approximately} offered useful extension to that work; however, the purpose of the optimization for the latter studies is quite different from ours. What they optimized is the selection of the \textit{type} of activity or question in a concept-learning task, which can be viewed as a stationary problem (the value of a teaching activity can be assumed to be constant).

\subsection{Our technique}

In this paper, we propose an approach that builds on the literature described above and draws its contributions together in a single system, with
(i) \textbf{model-based} estimation of the learner's memory state that accounts for forgetting, (ii) \textbf{online inference} of the model's parameters at \textbf{item level} for single users, and (iii) \textbf{online planning} that accounts for the user's practice schedule.
We evaluated the unified system in a controlled experiment, comparing it against a popular rule-based approach (involving the Leitner system).

\section{Individualized Model-Based Planning}
\subsection{General characterization of the problem}

We consider two agents interacting with each other: the \textbf{learner} and the \textbf{teacher}.
The goal of the learner is to learn a finite set of items $\mathcal{A}=\{a^1, a^2, \ldots, a^Q\}$ of length $Q$, and the teacher offers assistance by presenting the items to memorize.
In this paper, we address the problem of the teacher, which is to pick item $a \in \mathcal{A}$ that will help the learner to progress.

We assume that time is discrete.
The timing of the teaching sessions is controlled by the learner.
If the teaching process is not active at time $t$, the teacher cannot take any action: teaching is \textit{paused}, which we express as $a_t = 0$.
Otherwise, the learner is in a \textit{teaching session}: the teacher suggests item $a_t \in \mathcal{A}$ and observes $\omega_t \in \lbrace 0, 1 \rbrace$, indicating whether the learner knows item $a_t$ at this point in time.
In a practical example involving second-language vocabulary (e.g., German for an English-speaker), an item $a \in \mathcal{A}$ would be a word in the source and target languages (e.g., ``dog'' $\rightarrow$ ``Hund''); the source word (``dog'') is presented to the learner, who tries to respond with the target word (``Hund'').
We use $\omega_t$ to indicate whether the response was correct ($\omega_t = 1$) or not ($\omega_t = 0$).

Aiming to increase the learner's knowledge, the teacher sets learning objectives.
In practice, this entails defining a final reward for the teacher to receive after all the teaching sessions are complete.
The objective of the teacher is to maximize this reward, which depends on the actual memorization by the learner.

Were the learner to have unbounded memory abilities, the optimal sequence of items would be to present each item once.
In reality, human memorization is imperfect: the probability of recall of a previously seen item can be far lower than~1. Also, the learner's memory state is not known. The teacher must, therefore, rely on a memory model, which serves as a proxy for the learner's actual memorization state.
This model $M$ is parameterized by $\theta$ in parameter space $\Theta^M$. The model defines a probability $p_M(\omega_t | a_{t}, H_{t-1}, \theta)$, for the \textit{probability of recall}, with $H_{t-1} = (a_1, \dotsc, a_{t-1})$ being the history up to time $t-1$. The probability of recall denotes the probability of item $a \in \mathcal{A}$ being retained by the learner under  memory model $M$.
The teacher relies on a proxy reward $R(M, \theta, H_t)$ at time $t$ as an approximation for the actual reward. The optimization problem to be solved by the teacher is
\begin{equation}
    \text{arg}\max_{a_1, \dotsc, a_F \in \mathcal{A}} \quad R(M, \theta, H_F = (a_1, \dotsc, a_F))
    \label{eqn:objective}
\end{equation}
where $F$ is the end date of the teaching.

\subsection{POMDP formulation}

For solving Equation~\ref{eqn:objective}, our proposed technique is to model the overall teaching process for the teacher as a POMDP $(S, A, R, T, \Omega, O)$, where $S$ is the set of possible states and $A$ the set of actions, $R$ is a payoff function, $T$ is the state-transition probability, $\Omega$ is the set of observations, and $O: A \times S \times \Omega \rightarrow [0, 1]$ defines the probability $p(\omega | a, s) = O(a, s, \omega)$.
The actions correspond to presenting an item, so $A = \mathcal{A}$; the observations correspond to a success or a failure of the learner, so
$\Omega = \lbrace 0, 1 \rbrace$.
The state $s \in S$, describing the status of the memory (the probability of recall for each of the items, the values of the memory parameters, and so on), depends on the chosen memory model $M$. The payoff function always evaluates to 0, except in the very last step, for~$F$, for which the payoff is defined as the reward $R(M, \theta, H_F)$, where $H$ is the history of actions up until $F$. We set the state $s = (H, \theta)$ to be the Cartesian product of the history of actions $H$ and of the model parameterization $\theta$. Because the memory states $s$ are not directly observable, the teacher's problem at each interaction can be divided into two parts: (i) inference of the parameterization $\theta \in \Theta^M$ for the learner and (ii) item selection.
Therefore, we cut the POMDP into two complementary components, called the \textbf{psychologist} and \textbf{planner}, with the \textbf{learner} here being considered to be the environment.
The psychologist is endowed with a memory model $M$ and
is in charge of proceeding from the observations $\omega_t \in \Omega$
to infer its correct parameterization $\theta \in \Theta_M$, so as to estimate the recall probabilities. It accounts for the POMDP components $(T, O)$.
The planner, in turn, is endowed with a definition of learning that takes the form of the reward function $R$ and is in charge of, for each time point, selecting the optimal item to present while accounting for the future (see Figure~\ref{fig:teaser}).

\section{Implementation}
\subsection{The psychologist: The model and inference}

The psychologist is in charge of managing the memory model,~$M$ --- more specifically, with inferring its parameterization~$\theta \in \Theta$ and $p(\omega | a, H)$, the probability of recall of item~$a$ in light of history~$H$.

\paragraph{Memory models with exponential forgetting.} The \textbf{exponential forgetting} (EF) model is based on Ebbinghaus's model~\cite{ebbinghaus1885uber} and in this respect is close to the one used by Settles and Meeder~\shortcite{settles2016trainable}, Tabibian et al.~\shortcite{tabibian2019enhancing}, or Hunziker et al.~\cite{hunziker2019teaching}. This model provides both a description of the \textbf{lag} effect (the shorter the time since the last review, the higher the probability of recall) and the \textbf{repetition} effect (the more repetitions there are, the higher the probability of recall). It assumes that the probability of recall decays exponentially but that the decay increases each time a given item is reviewed.
In this model, the probability of recall of an item $a$ is given by

\begin{equation}
    p_{EF}(\omega_t = 1 \mid a_t, H_{t-1}, \theta) = e^{- \alpha (1 - \beta)^{n(a_t; H) - 1} \Delta t_{a_t}(H_{t-1};t)}
    \label{eqn:recall-ef}
\end{equation}

where $\Delta t_{a}(H_{t-1};t)$ represents the time that has elapsed since the last presentation of the item $a$ and $n(a;H)$ is the number of times that said item has been presented in history $H$.
The parameters $\alpha \in [0, +\infty)$ and $\beta \in (0,1)$ correspond to the initial forgetting rate and the effect of teaching, respectively. The parameter space for the exponential forgetting model, then, is $\Theta_{EF} = \lbrace (\alpha, \beta): \alpha \in [0, +\infty), \beta \in (0,1) \rbrace$.
Importantly, $\alpha$ and $\beta$ are specific to each learner and have to be inferred from the teaching interactions.

\paragraph{Memory models with item-specific exponential forgetting.} We introduce a variant of the exponential forgetting model, called \textbf{item-specific exponential forgetting} (ISEF).
This model diverges from the EF model in that the parameters $(\alpha, \beta)$ are not global for a specific user but defined for each item separately.
The corresponding parameter space $\Theta_{ISEF}$ is then $\Theta_{ISEF} = \Theta_{EF}^{|\mathcal{A}|}$, where $|\mathcal{A}|$ is the number of items to learn.
The number of parameters to be estimated by the psychologist in this model is proportional, then, to the number of elements for learning.

\paragraph{Inference methods: Bayesian belief updating.} In order to compute the recall probabilities, the psychologist needs to possess accurate knowledge of the learner, which includes the memory model (discussed above) and its parameter $\theta$.
This parameter can be inferred from observation of the interactions with the learner.

At each interaction point, we assume that the teacher proceeds with Bayesian belief updating for the parameter~$\theta$.
If the item $a_t \in A$ is presented and the outcome $\omega_t\in \lbrace 0, 1 \rbrace$ is observed, the Bayesian belief update obeys the following rule:

\begin{equation}
     p_M^{t+1}(\theta) = \dfrac{
      p_M^t(\omega_t \mid a_t, H_{t-1} \theta) p_M^t(\theta)
     }{
        \int p_M^t(\omega_t  \mid a_t, H_{t-1}, \theta) p_M^t(\theta) d\theta
     }
     \label{eqn:belief-update}
\end{equation}
Here, $p_M^{t+1}(\theta)$ is the posterior probability of the parameter $\theta \in \Theta$ and $p_M^t(\theta)$ the prior belief about this parameter. % Clear enough? -als
In this Bayesian belief update, the likelihood corresponds to the probability of recall, which depends on both the model $M$ and the parameter $\theta$.

\subsection{The planner: Reward and planning algorithms}

The planner is in charge of selecting the optimal items to present with regard to achieving some fixed learning objective.

\paragraph{The reward: Counting the items learned.} A simple learning objective for the teacher is to maximize the number of items learned in the course of the teaching process.
In human experience, this would correspond to grading a final test in which the learner is asked about all the items presented earlier. Since the result of such a test cannot be known at planning time, we use the following definition as a proxy: an item $a \in \mathcal{A}$ is considered to be known at level $\rho \in [0,1]$ for a memory model~$(M, \theta)$ if $p_M(\omega_t = 1 | a_t = a, H_{t-1}, \theta) \geq \rho$.
Operating with this definition, we define the final reward as the number of items known at level $\rho$ after the final teaching interaction at step $F > 0$:
\begin{equation}
R_{\rho}(M, \theta, H_F) =  \left| \lbrace a \in \mathcal{A} : p_M(\omega_{F+1} = 1 | a_{F+1} = a, H_{F}, \theta) \geq \rho \rbrace \right|
\label{eqn:reward}
\end{equation}
where $|.|$ denotes the number of elements in a set.

\paragraph{Planning algorithms: Myopic sampling.} The myopic sampling algorithm attempts to maximize the final reward by, at each time step $t$, choosing the action that maximizes the immediate reward $R(M, \theta, H_t)$.
This planning is greedy and does not require fixing the horizon $F$.

For the item-counting reward (see Equation~\ref{eqn:reward}), myopic sampling induces straightforward behavior.
Two cases are possible in choosing an item: If at least one item has a probability of recall below $\rho$, one can increment the immediate reward by 1 by presenting one of the ``forgotten'' items to the learner.
Otherwise, the algorithm presents a new item, since displaying a previously seen one would not affect the immediate reward.

\paragraph{Planning algorithms: Conservative sampling.} The main limitation of myopic sampling is that it does not do any planning. The decision tree expands exponentially (size = $N^t$), and the decision problem is dynamic (e.g., presenting $i$ at $t$ influences the effect of presenting $i$ at $t+1$). Using a brute-force algorithm to explore each solution is impossible. Since such standard sampling techniques as Monte Carlo tree search are computationally demanding in our online setting, we offer a variant algorithm, \textbf{conservative sampling}, as a solution.

The intuition behind the conservative-sampling algorithm is that we should \textit{avoid} presenting a particular item when there is insufficient time for learning that item or when this choice would adversely affect learning of any item introduced earlier.
More precisely, the algorithm adapts myopic sampling thus: using a myopic planner with a restricted number of items to introduce, our technique evaluates whether presenting the chosen item is compatible with memorizing the set containing all those previously seen by the learner. Specifically, this technique requires fixing the horizon $F$. Given the choice of item $a^i$, we consider the set of items $\mathcal{A}_{<a} = \lbrace a^1, \dotsc, a^i \rbrace$ that have been introduced before $a$'s first presentation (by convention, $a^j$ refers to the $j$th new item introduced in the history). If the item considered is $a^1$, then $a^1$ is selected. Otherwise, our criterion for presenting item $a=a^i$ is that the set $\mathcal{A}_{<a}$ can still be fully memorized by the learner at $F$ under teaching by a myopic teacher.
When this criterion is not satisfied, we use the myopic sampling algorithm to select another item considering as possibilities only the set $\lbrace a^1, \dotsc, a^{i-1} \rbrace$ and we repeat the process.

\paragraph{Planning constraints: Learning with breaks.} Our framework also introduces the possibility of considering breaks in the implementation of the planning phase. In general, breaks can be stochastic or deterministic.
For simplicity's sake, here we consider them deterministic, and known by the teacher.

\section{The Simulation Studies}
% We now describe the experiments with artificial learners.
\subsection{The baseline: The Leitner system}

We chose a Leitner system as the baseline, since this affords comparison with a popular adaptive approach.
This system is unlike model-based ones in that it is competitive in terms of adaptation to the learner.
The general principle of a Leitner system is the following: in the event of success ($\omega_t = 1$), lower the frequency of review; in the event of failure ($\omega_t = 0$), increase the review frequency.

A classic implementation of this principle makes use of a ``box'' system: Each item $a \in A$ belongs to a box $k\in \mathbb{N}$.
The first time an item is presented, it is added to box $1$. During subsequent reviews, the item is moved from box $k$ to either box $k+1$ (in cases of success) or box $\max \{0, k-1\}$ (otherwise). The next time an item is presented for review depends on the box that item belongs to and is equal to $\delta_{A} \cdot \delta_{B}^{k}$, where $\delta_{A}$ and $\delta_{B}$ are scale parameters that influence how the reviews are spaced.

At each iteration $t$, only one item can be selected, even though it may be time to review several items under the aforementioned criterion.
An item that was not selected at $t$ even though $t$ was equal to (or past) its time for review is added to a waiting queue.
The selection process is performed thus: (i) select an item that needs to be reviewed in accordance with the first criterion; (ii) if several items qualify, select the one that has spent the most iteration in the waiting queue; (iii) in the event of equality, select the one with the smallest $k$ value; and (iv) if the figures are still the same, select randomly.
If no item meets the first criterion, a new item gets introduced.

\subsection{Methods}

\paragraph{Procedure.} Under our approach, we intend to simulate teaching in a realistic context.
Consequently, we simulate conditions that assume a few minutes of practice every day for a week, with an evaluation at the end of the week.

The artificial learners use the exponential forgetting model.
Our tests applied two variants of the model:
a \textbf{non-item-specific} one, wherein each user has a specific parameterization but all the items from any given user share the same parameterization (for proof of concept), and an \textbf{item-specific} one, in which, for each user, each item has a specific parameterization (offering finer granularity). Also, to demonstrate the influence of the inference process on the results (including any interference exhibited), we contrasted two degrees of knowledge held by the psychologist: an \textbf{omniscient} psychologist has access to the parameterization of the items/learners, and a \textbf{non-omniscient} psychologist does not.

\paragraph{Execution.} The simulations ran on a cluster hosted by Aalto University, the technical characteristics of which are accessible at \url{https://scicomp.aalto.fi/triton/overview/}. The code for reproducing the simulations, our analysis, and figures are available at \url{https://github.com/AurelienNioche/ActiveTeachingModel/tree/master}.

\paragraph{Parameterization.} $N=100$ artificial agents were simulated for $6$ training sessions, with a one-day break between sessions.
The number of items learned was assessed in an evaluation session on the seventh day.
An item was considered learned if its probability of recall was above the threshold $\rho=0.90$.
The total number of items was set to $Q=500$, and
the time for completion of one iteration was set to $4$~s.

\begin{figure}[!ht]
    \centering
    \includegraphics[width=\columnwidth]{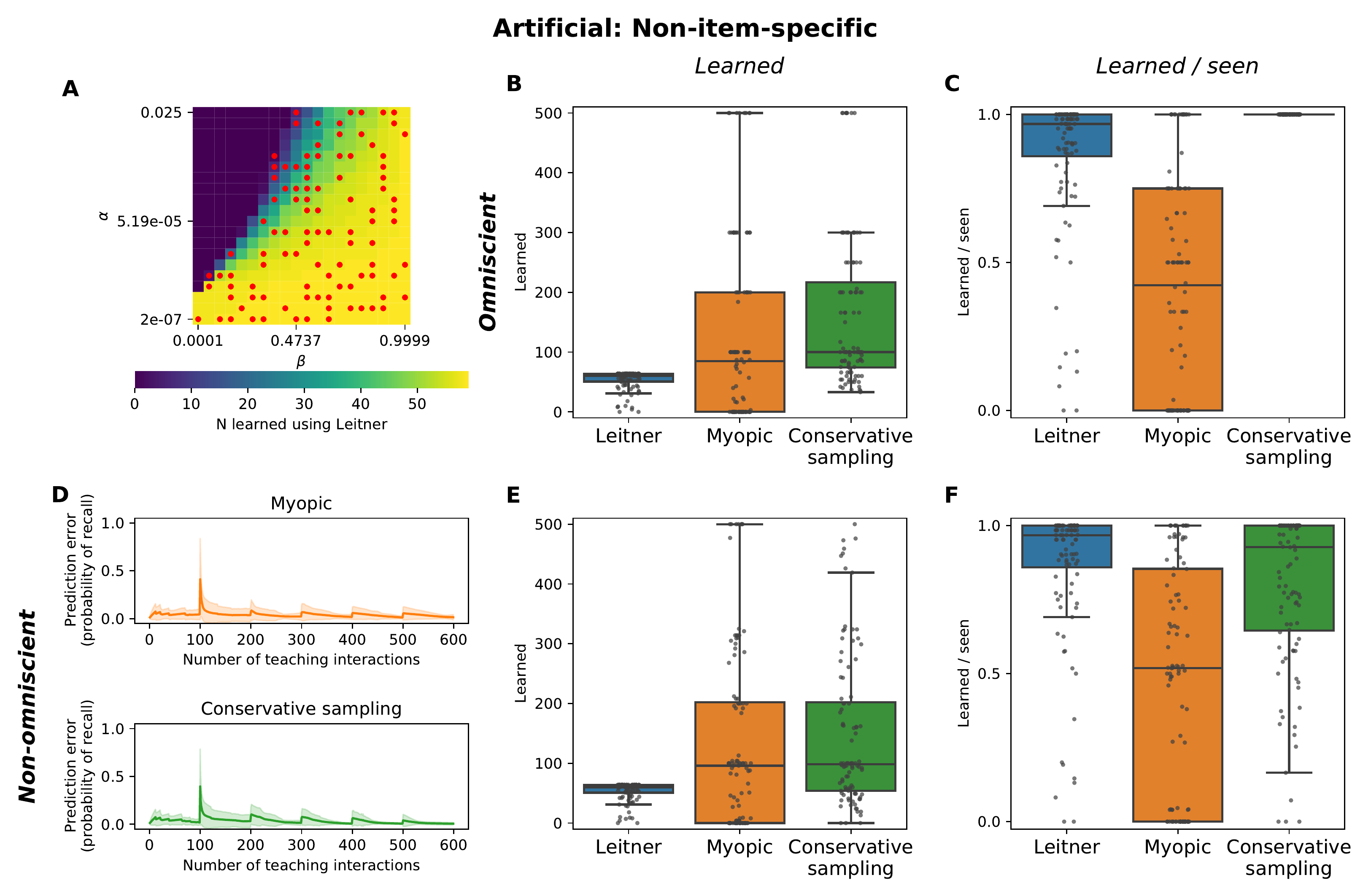}
    \caption{Artificial learners' performance in the \textbf{non-item-specific} condition. One dot represents one learner.
    Pane A\@: Parameters used to create the learners (the color gradient refers to the number of items learned in the Leitner system).
    B\@: Items learned with an \textbf{omniscient} psychologist.
    C\@: With an \textbf{omniscient} psychologist, the ratio between the items learned and all items seen.
    D\@: Average prediction error (shading denotes the area within one SD).
    E\@: Items learned with a \textbf{non-omniscient} psychologist.
    F\@: With a \textbf{non-omniscient} psychologist, the learned/seen-item ratio.
    }\label{fig:artificial_nospec}
\end{figure}

% Learners
For the artificial learners, we chose a parameterization that is plausible for humans by running $400$ exploratory simulations. In each simulation, we supplied the agent with parameters $(\alpha, \beta)$ picked from the Cartesian product of 20 values for each parameter, with $\alpha \in [2\mathrm{e}{-07}, 0.025]$ and $\beta \in [0.0001, 0.9999]$, using a logarithmic scale for the value of $\alpha$. Note that the bounds for the forgetting rate were within the limits of what we could observe in the user study (even if unlikely) for the time of around one iteration (with $\alpha = 0.025$, the probability of recall falling under $0.90$ in four seconds) and around one week (with $\alpha = 2\mathrm{e}{-07}$). The corresponding bounds for $\beta$, in turn, represent a repetition effect that is practically negligible (when $\beta = 0001$, the forgetting rate is nearly unchanged after a new repetition) and to a forgetting rate reduced by almost $100\%$ (when $\beta = 0.9999$).
On the basis of these preliminary simulations, we chose parameters for the next simulations -- the ones used for our analysis --- such that they equated to agents learning at least one item with a Leitner system (see Figure~\ref{fig:artificial_nospec}).

% Psychologist
For the psychologist component of our system, the parameters are evaluated with a grid of size $100 \times 100$ ($\alpha$: log scale bounded by $[2\mathrm{e}{-7}, 2.5\mathrm{e}{-2}]$; $\beta$: linear scale bounded by $[0.0001, 0.9999]$).
 The prior over the parameter values is uniform at initialization.
In the \textit{item-specific} condition, once at least one item has been reviewed, the prior for an item not reviewed is the average of the priors of the items reviewed at least once.
For the conservative-sampling algorithm, the estimated time of completion of one iteration is $4$~s.

For the Leitner system, $\delta_{A}=4$, $\delta_{B}=2$, which means that items in box $0$ should be reviewed after four seconds, items in box $1$ eight seconds later, then $16$ seconds, $32$ seconds, etc.

\paragraph{Comparisons.}
Our comparisons examined both (i) the number of items learned (an item is regarded as \textit{learned} if its probability of recall exceeds $\rho$ at the time of the evaluation session) and (ii) the ratio between the number of items learned and that of all items seen, across conditions: a teacher using a Leitner algorithm (a ``Leitner teacher''), myopic sampling (a ``myopic teacher''), or a conservative-sampling algorithm (a ``conservative-sampling teacher''), in both conditions of knowledge on the part of the psychologist (omniscient and not omniscient).
The latter ratio was assessed to identify potential teaching strategies, relying mainly on wide ranges of items, non-selectivity, and overwhelming ranges of item presentation.
We evaluated statistical significance via a non-parametric test (Mann-Whitney $U$, with significance threshold $p=0.05$). Bonferroni corrections for multiple comparisons were applied (since two comparisons were made for each dataset, the $p$-values are multiplied by $2$).

Each box plot extends from the lowest to the highest quartile of the frequencies observed (the central line denotes the median, and the whiskers refer to an IQR of 1.5).

% Figure 2: Artificial nospec

\subsection{Results}
% Nspec-omni
In the
\textbf{non-item-specific} condition with an \textbf{omniscient} psychologist (see Figure~\ref{fig:artificial_nospec}), the simplest scenario under our framework, the number of items learned does not differ between the myopic (M) and Leitner (L) teachers ($u=5650.0$, $p=0.110$, $p_{cor}=0.221$, $N=100\times 2$), while the conservative-sampling (CS) teacher significantly outperforms the Leitner one ($u=8539.0$, $p<0.001$, $p_{cor}<0.001$, $N=100\times 2$).
The ratio between the items learned and all items seen is significantly different from the baseline value for both model-based teachers (M vs. L\@: $u=1836.5$, $p<0.001$, $p_{cor}<0.001$, $N=100\times 2$; CS vs. L\@: $u=8250.0$, $p<0.001$, $p_{cor}<0.001$, $N=100\times 2$): the myopic teacher leaves more items \textit{unlearned} among those seen (at least once) by the learner than the baseline teacher, while the CS teacher leaves fewer items \textit{unlearned} than the baseline.

% Nspec-Nomni
In the \textbf{non-item-specific} condition with a \textbf{non-omniscient} psychologist (see Figure~\ref{fig:artificial_nospec}), the number of items learned is significantly greater for each model-based teacher relative to the baseline (M vs. L\@: $u=5976.5$, $p=0.017$, $p_{cor}=0.033$, $N=100 \times 2$; CS vs. L\@: $u=7401.5$, $p<0.001$, $p_{cor}<0.001$, $N=100 \times 2$). As with an \textbf{omniscient} psychologist, the ratio of items learned to items seen is significantly lower than the baseline level for the myopic teacher ($u=2065.0$, $p<0.001$, $p_{cor}<0.001$, $N=100\times 2$).
This time, no significant difference is visible for the CS teacher ($u=4551.5$, $p=0.261$, $p_{cor}=0.521$, $N=100\times 2$). For both model-based teachers, the prediction error decreases with time (it starts at 0 since the scenario begins with new items always being supplied alongside the correct answer, and the ``spikes'' correspond to the beginning of each session, where the space between two sessions induces an increase in the magnitude of the forgetting).

% Figure 3: Artificial spec
\begin{figure}[!ht]
    \centering
    \includegraphics[width=\columnwidth]{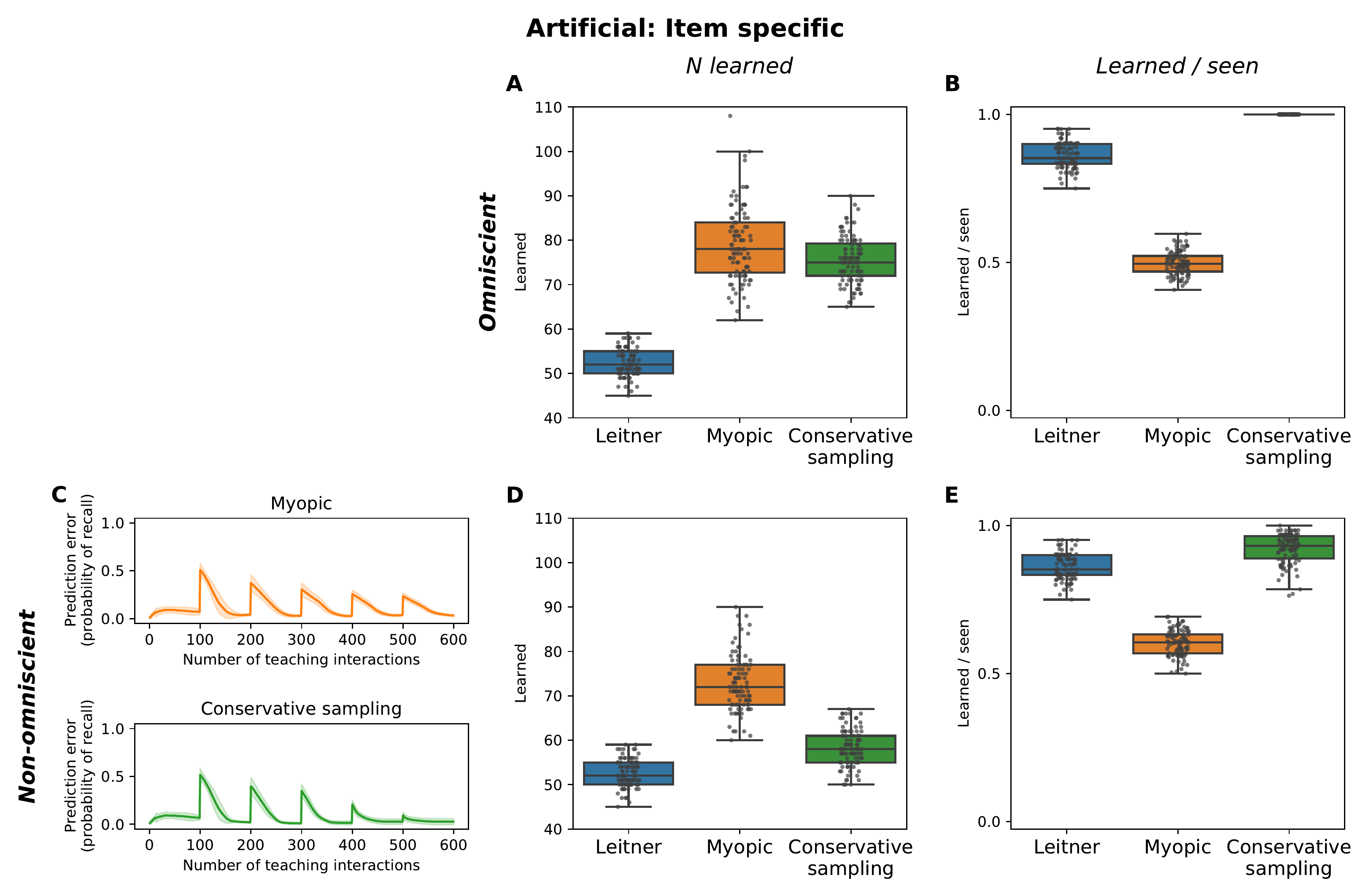}
    \caption{Artificial agents' performance in the \textbf{item-specific} condition. One dot represents one learner.
    A\@: Items learned when the psychologist is omniscient.
    B\@: The ratio between items learned and all items seen when the psychologist is \textbf{omniscient}.
    C\@: Average prediction error (shading denotes the one-SD band).
    D\@: Items learned when the psychologist is \textbf{not omniscient}.
    E\@: The ratio of items learned to items seen when the psychologist is \textbf{not omniscient}.
    }\label{fig:artificial_spec}
\end{figure}

% spec-omni
In the \textbf{item-specific} condition with an \textbf{omniscient} psychologist (see Figure~\ref{fig:artificial_spec}), the number of items learned is significantly greater for both model-based teachers as compared with the Leitner teacher (M vs. L\@: $u=10,000$, $p<0.001$, $p_{cor}<0.001$, $N=100\times 2$; CS vs. L\@: $u=10,000$, $p<0.001$, $p_{cor}<0.001$, $N=100\times 2$).
The learned-to-seen ratio is significantly lower with the myopic and Leitner teachers ($u=0$, $p<0.001$, $p_{cor}<0.001$, $N=100\times 2$), while it is higher for the conservative-sampling teacher ($u=10,000$, $p<0.001$, $p_{cor}<0.001$, $N=100\times 2$).

% spec-Nomni
When the framework is evaluated in the maximum-complexity condition --- i.e., in the \textbf{item-specific} condition with a \textbf{non-omniscient} psychologist --- the number of items learned is significantly greater for both the myopic and the conservative-sampling teacher relative to the Leitner teacher (M vs. L\@: $u=10,000$, $p<0.001$, $p_{cor}<0.001$, $N=100\times 2$; CS vs. L\@: $u=8640$, $p<0.001$, $p_{cor}<0.001$, $N=100\times 2$). As for the ratio of items learned to items seen, the myopic teacher displays weaker performance than the baseline teacher ($u=0$, $p<0.001$, $p_{cor}<0.001$, $N=100\times 2$), while the CS teacher significantly outperforms it ($u=8294.5$, $p<0.001$, $p_{cor}<0.001$, $N=100\times 2$). For both model-based teachers, the error of prediction diminishes over time as expected.

\section{The User Study}
\subsection{Methods}

\paragraph{Participants.} After having received ethics approval for the study from Aalto University per the university's guidelines, we recruited 65 individuals through a mailing list of Aalto University, of whom 53 completed the task (self-reported data: 12 male, 39 female, and 1 other; $\text{age}=26.38$ years, SD $\pm 7.67$).
All participants provided informed consent before proceeding to the experiment itself. This included acknowledgment of their right to withdraw from the study at any time without fear of negative consequences.
At this stage, the participants also filled in a survey form asking their age, gender, mother tongue/s, and secondary languages.
They were compensated for their time with cinema vouchers.

\paragraph{Procedure.} The study was conducted online. Participants connected to a Web-based application displaying a graphical user interface similar to that of other flashcard software (see Figure~\ref{fig:teaser}). We used a mixed experimental design.
Each participant engaged with two distinct teachers: (i) a Leitner teacher (the baseline, identical to that in the experiments with artificial learners) and (ii) a model-based planning system. For each user, the model-based planning system used the ISEF memory model, but the planning was based either on a myopic sampling algorithm or on a conservative-sampling algorithm.
Each participant was assigned two sets of $Q = 200$ items randomly drawn from a database containing approx. $2,000$ pairs of kanji-English definitions (one set for each teacher, for $400$ items in total).
There were two (roughly 10-minute) training sessions, with 100 questions each, per day for six days, then two evaluation sessions on the seventh day.
Each question consisted of the presentation of a kanji and six possible answers (note that the six-responses setting corresponds to not too high a probability of randomly selecting the correct answer while still not overloading the interface).

The first time any given kanji was presented, the user was shown the right answer. Each set of two sessions took place at the same time of the day, chosen by each participant at the time of signing up for the experiment, and kept constant for all seven consecutive days.
For one of the series, the teacher handling the item selection was a Leitner teacher, and the other session's teacher was a version of our system (either myopic or conservative sampling).
The order of the series was alternated between the two teachers: if the first session on one day used teacher $A$, the next day's first session was the one with teacher $B$.

\paragraph{Implementation and execution.} The software used for the experimental part of the study is based on a client/server architecture. The client part was developed on the Unity platform and ran in the subjects' Web browser by means of the WebGL API\@. The server part was developed via the Django framework. The code for the server portion is available at \url{https://github.com/AurelienNioche/ActiveTeachingServer/tree/master}, and
the Unity assets are available at \url{https://github.com/AurelienNioche/ActiveTeachingUnityAssets/tree/master}. The application was hosted on an Aalto University virtual server (Ubuntu 18.04 $\times$86\_64 GNU/Linux 4.15.0--91-generic, with 4~GB of RAM, two virtual CPU cores, and 50~GB of disk space).

\paragraph{Parameterization.} The parameterization is similar to that for the artificial learners in the \textbf{item-specific} condition with a \textbf{non-omniscient} psychologist except that (i) the number of items to learn per teacher was 200 (instead of 500), to avoid very long evaluation sessions, and (ii) the space between two sessions managed by the same teacher was only approximately (rather than strictly) equal to one day, on account of the alternation between the two teachers.

\paragraph{Comparisons.} We compared both (i) the number of items learned (since probabilities of recall were not directly accessible, an item was considered \textit{learned} if it was successfully recalled twice during the evaluation session) and (ii) the ratio between of items learned to all items seen between a Leitner teacher and the model-based teacher (a myopic or CS teacher, depending on the subject). The statistical procedure was identical to the one employed for the corresponding comparison with artificial agents.

\subsection{Results}

% Figure 4: Human
\begin{figure}[!ht]
    \centering
    \includegraphics[width=\columnwidth]{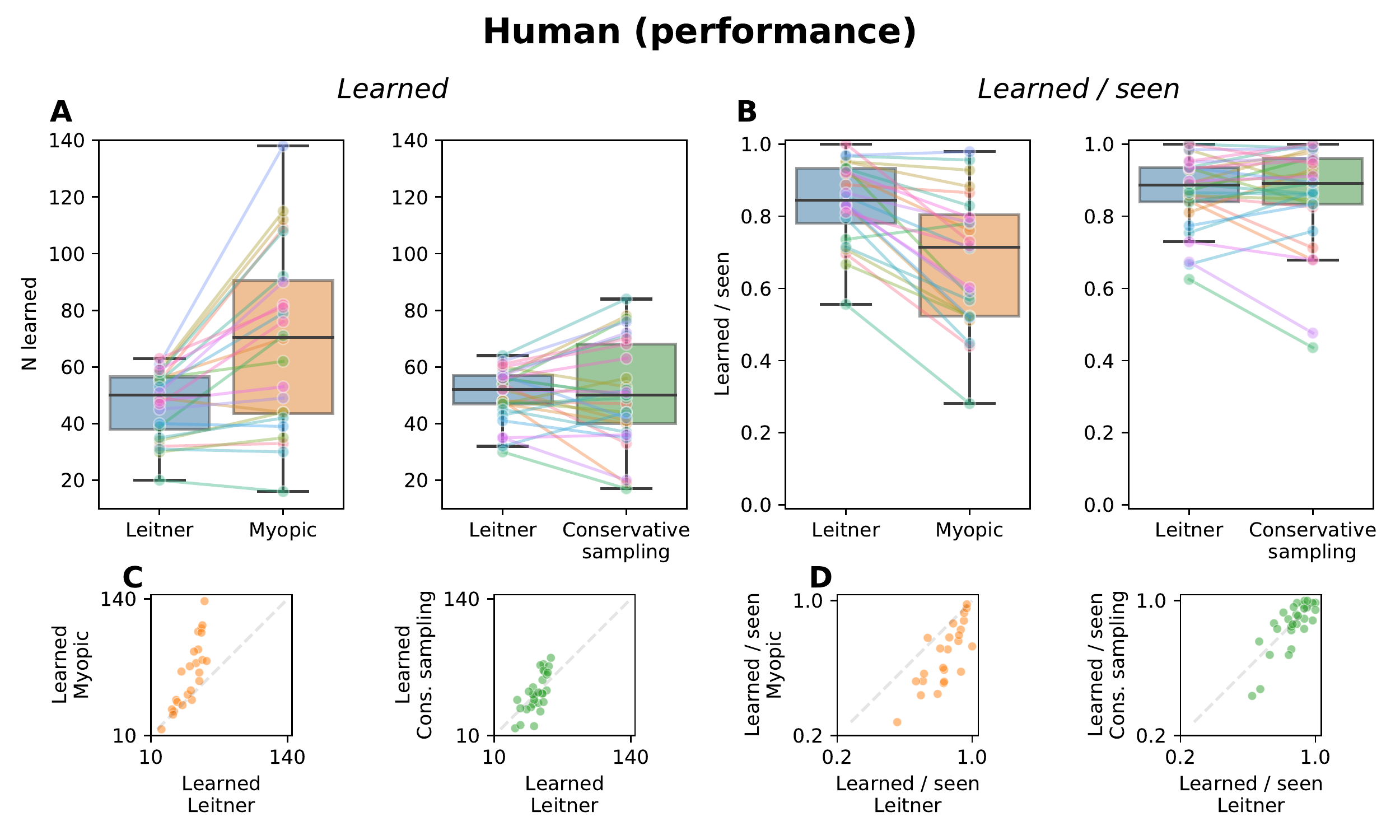}
    \caption{Human learners' performance.
    Each dot represents one learner (M vs. L\@: $N=24$, CS vs. L\@: $N=29$).
    Panes A and C present the items learned, and
    panes B and D show the ratio of items learned to items seen.
    }\label{fig:human}
\end{figure}

With human learners (see Figure~\ref{fig:human}), the number of items learned proved to be significantly greater with the myopic teacher than with the Leitner teacher ($u=174.5$, $p=0.019$, $p_{cor}=0.038$, $N=24\times 2$). In contrast against the corresponding results obtained with artificial learners, no significant difference was visible between the CS and the Leitner teacher ($u=434.5$, $p=0.828$, $p_{cor}=1.655$, $N=29\times 2$). In a parallel with the artificial agents in the \textbf{item-specific} condition with a \textbf{non-omniscient} psychologist, the ratio of items learned to items seen for the myopic teacher was below the baseline figure ($u=435$, $p=0.003$, $p_{cor}=0.005$, $N=24\times 2$), while no significant difference emerged for the CS teacher ($u=397$, $p=0.721$, $p_{cor}=1.441$, $N=29\times 2$).

\begin{figure}[!ht]
    \centering
    \includegraphics[width=\columnwidth]{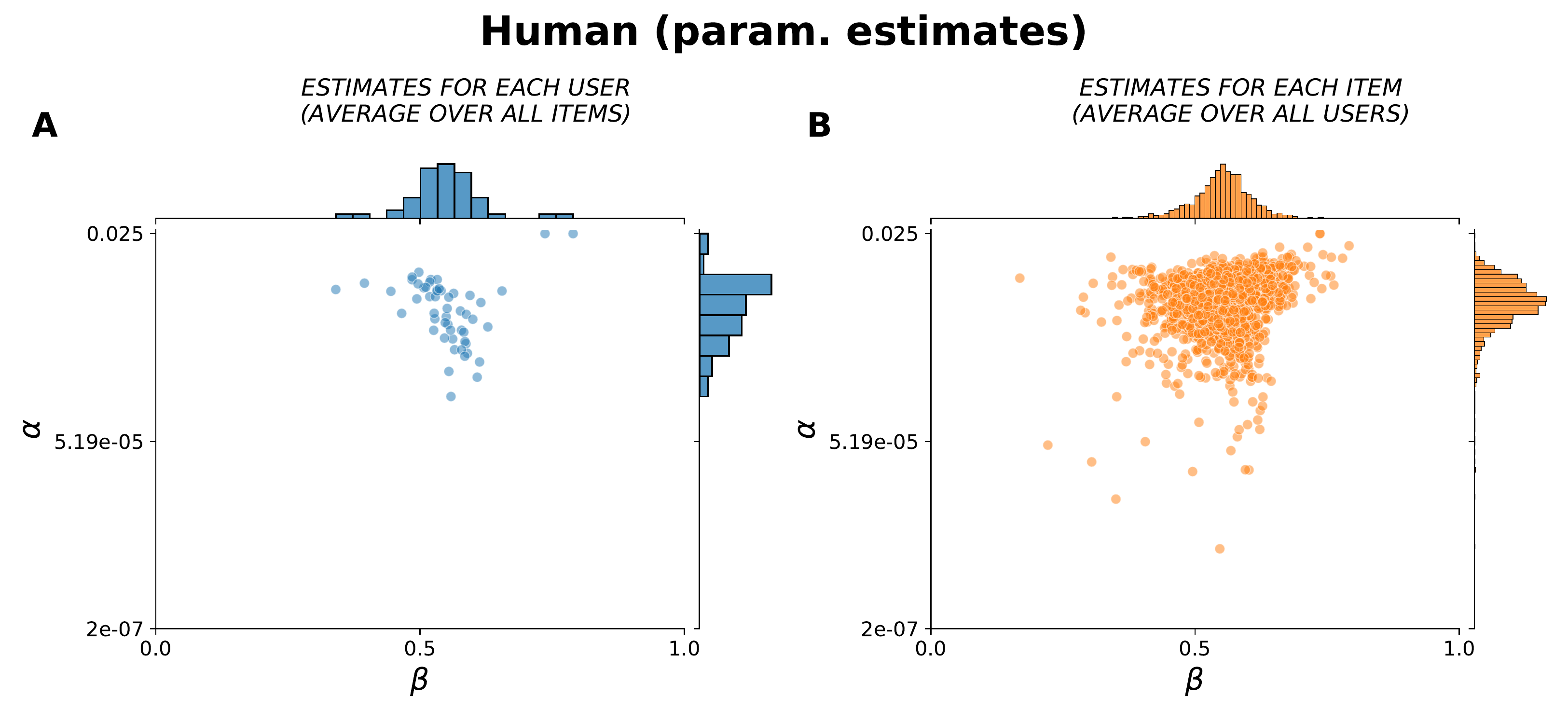}
    \caption{Parameter estimates for human learners.
    A\@: Estimates for each user (with averaging over all items), where each dot represents one learner ($N=53$).
    B\@: Estimates for each item (averaged over all users), where each dot represents one item ($N=2047$).
    }\label{fig:estimates}
\end{figure}

% In order to highlight the importance of individualization,
To verify that the teacher actually adapted at the level of users and items, we present the final parameter estimates made by the psychologist in Figure~\ref{fig:estimates}.

% Figure 5: Estimates

\section{Discussion and Conclusions}
Our work extends research on model-based approaches for tutoring systems by offering a modular framework to combine \textbf{online inference} specific to \textbf{each user} and \textbf{each item} with \textbf{online planning} that takes the learner's time constraints into account. As presented above, an implementation of this framework with a simple memory model showed favorable results; hence, it warrants further research. Our experiments, with both artificial and human learners, attest that even a simple model can offer performances at least equalling that of the industry-standard Leitner method, as evaluated by two metrics: ``$n$ learned'', which captures (raw) performance, and the ``$n$ learned / $n$ seen'' ratio, which can be interpreted as the ``precision'' of the teacher. Furthermore, we observed that our method successfully adapts to the variability of learners' learning capabilities and items' difficulty level.

In our simulations with a learner that, by construction, has exact correspondence with the model, the long-term-planning teacher did better than the Leitner and myopic teachers --- markedly and consistently. We were surprised to find, however, that with human learners the myopic teacher, which greedily picks the best next item to show, fared better than did the longer-term planning approach.
From our current knowledge, we attribute this finding to the predictive model in our framework, because the results with artificial learners were conclusive while the ones for humans were not.
Moreover, any errors in inference are bound to be compounded in planning, and this could explain why the myopic planner maintained its performance.
These observations highlight the importance of developing more robust planning methods in future work.

The main contribution represented by this paper has less to do with the performance of an end-to-end method than it does with proposing a modular framework that, in contrast to prior methods, clearly advances our tackling of multiple aspects of learning. For instance, it may be possible to include elements such as additional cognitive aspects of memorization, among them subtler ones (such as the shape of the forgetting curve~\cite{rubin1996one,heathcote2000power} and modeling the spacing effect separately~\cite{cepeda2006distributed,walsh2018evaluating}). Among other factors for possible consideration are the influence of words' similarity and/or of context on memorization \begin{CJK}{UTF8}{min} (e.g.\ how learning ``八'': ``eight'' interferes with learning ``人'': ``people'')\end{CJK}, non-binary responses, and irregular learning sessions (for which one might employ a predictive model of the learner's ``favorite'' moments in the course of the day, with the teacher's expected rewards getting weighted accordingly). Also, extensions are possible for addressing the nature or type of various teaching objectives: focusing on particular classes of words, paying attention to exam dates, attention to quality over quantity, etc. Advances in all of these directions can be easily integrated into our framework, and attention to all the various aspects listed is necessary for enhancing the learning experience of real human users. Finally, we would expect this framework to hold great potential for extension to other application contexts wherein adaptation to the user and other individualization would be beneficial.

%% Acknowledgements
\begin{acks}
We thank all study participants for their time, and our colleagues and the reviewers for their helpful comments. This work was funded by Aalto University's Department of Communications and Networking (Comnet), the Finnish Center for Artificial Intelligence (FCAI), the Foundation for Aalto University Science and Technology, and the Academy of Finland (projects 328813, ``Human Automata,'' and 318559, ``BAD''). We also acknowledge the computation resources provided by the university's Science-IT project.
\end{acks}

\bibliographystyle{ACM-Reference-Format}
\bibliography{main}

\appendix

\end{document}